\documentclass[aps,pre,twocolumn,superscriptaddress,showpacs,preprintnumbers,amsmath,amssymb]{revtex4-1}

\usepackage{graphicx}
\usepackage{multirow}
\usepackage[usenames]{color}

\usepackage{hyperref}

\DeclareMathOperator{\Real}{Re}

\DeclareMathOperator{\sgn}{sgn}

\usepackage{color,soul}
\setul{0.5ex}{0.5ex}
\setulcolor{red}

\begin{document}

\title{Superconducting spin-valve effect in heterostructures with ferromagnetic Heusler alloy layers}

\author{A.~A.~Kamashev}
\affiliation{Zavoisky Physical-Technical Institute, FRC Kazan
Scientific Center of RAS, 420029 Kazan, Russia}

\author{N.~N.~Garif'yanov}
\affiliation{Zavoisky Physical-Technical Institute, FRC Kazan
Scientific Center of RAS, 420029 Kazan, Russia}

\author{A.~A.~Validov}
\affiliation{Zavoisky Physical-Technical Institute, FRC Kazan
Scientific Center of RAS, 420029 Kazan, Russia}

\author{J.~Schumann}
\affiliation{Leibniz Institute for Solid State and Materials
Research IFW Dresden, D-01069 Dresden, Germany}

\author{V.~Kataev}
\affiliation{Leibniz Institute for Solid State and Materials
Research IFW Dresden, D-01069 Dresden, Germany}

\author{B.~B\"{u}chner}
\affiliation{Leibniz Institute for Solid State and Materials
Research IFW Dresden, D-01069 Dresden, Germany}
\affiliation{Institute for Solid State and Materials Physics, Technical University
Dresden, D-01069 Dresden, Germany}

\author{Ya.~V.~Fominov}
\affiliation{L.~D.\ Landau Institute for Theoretical Physics RAS, 142432 Chernogolovka, Russia}
\affiliation{Moscow Institute of Physics and Technology, 141700 Dolgoprudny, Russia}
\affiliation{National Research University Higher School of Economics, 101000 Moscow, Russia}

\author{I.~A.~Garifullin}
\affiliation{Zavoisky Physical-Technical Institute, FRC Kazan Scientific Center of RAS, 420029 Kazan, Russia}

\date{\today}

\begin{abstract}

We report a comparative analysis and theoretical description of the superconducting properties of two 
spin-valve-valve structures containing the Heusler alloy Co$_2$Cr$_{1-x}$Fe$_x$Al$_{y}$ 
as one of two ferromagnetic (F1 or F2) layers of the F1/F2/S structure, 
where S stands for the superconducting Pb layer. 
In our experiments we used the
Heusler alloy layer in two roles: as a weak
ferromagnet on the place of the F2 layer and as a half-metal on the
place of the F1 layer. In the first case, we obtained a large
ordinary superconducting spin-valve effect $\Delta T_c$ assisted by
the triplet superconducting spin-valve effect $\Delta T_c^{trip}$.
In the second case, we observed a giant magnitude of $\Delta
T_c^{trip}$ reaching 0.5\,K. An underlying theory based on the solution of the Usadel
equations using Kupriyanov-Lukichev boundary conditions with
arbitrary material parameters for all layers and arbitrary boundary
parameters for all interfaces is presented in Appendix. We find a good agreement between our experimental data and theoretical results.

\pacs{85.75.-d, 74.45+c, 74.25.Nf, 74.78.Fk}

\keywords{superconductor,ferromagnet,proximity effect}

\end{abstract}

\maketitle

\section{Introduction}

In the past two decades there has been an enormous theoretical and experimental interest to the development of the  elements for superconducting spintronics
(see, e.g., \cite{Ioffe,Feigelman}). Among those works, studies of the
superconductor/ferromagnet/superconductor (S/F/S) heterostructures
were considered to be promising for their use in the elements of quantum
logics \cite{Ryazanov1999}. An element of the quantum cubit
\cite{Ryazanov2001,Ryazanov2000} is based on the so-called Josephson
$\pi$-contact \cite{Ryazanov2001a,Kontos} which can be
realized in the S/F/S thin film multilayer. In this respect the S/F contact
has attracted a long-standing fundamental interest \cite{Ryazanov2004}.
In particular, the possibility to build up a superconducting (SC)
spin valve (SSV) based on the S/F proximity effect has been
theoretically suggested by Oh {\it et al.} \cite{Oh}. They proposed an
F1/F2/S trilayer with the S layer placed on the
top of the two F layers which magnetizations are decoupled. In this theory
the critical temperature $T_c$ of the SC heterostructure for the antiparallel (AP) mutual orientation of
magnetizations should be larger than for the parallel (P) one. This is
because the mean exchange field from two F layers acting on the
Cooper pairs in the S layer is smaller for the AP configuration as compared to the P case.  An alternative structure F/S/F with the similar operational principle was proposed theoretically by
Tagirov \cite{Tagirov}. However, it took more than ten
years before a full switching between the normal and the SC
states of such a device was realized experimentally \cite{Leksin2010}.

Importantly, Fominov {\it et al.} \cite{Fominov} have theoretically shown that the interference of the
Cooper pairs wave function reflected from both surfaces of the F2
layer in the F1/F2/S structure may be constructive or destructive. It depends on the thickness of the F layer, yielding the  direct and the inverse SSV effect, respectively.
Indeed, a thickness dependent sign-changing oscillating behavior of the SSV
was experimentally observed in Ref.~\cite{Leksin2011}.
Therefore, the initial assumption by Oh {\it et al.} \cite{Oh} that the
AP mutual orientation of magnetizations of the F1 and F2 layers is
preferable for superconductivity may
not be always correct.

\begin{figure}[h]
\center{\includegraphics[width=1\linewidth]{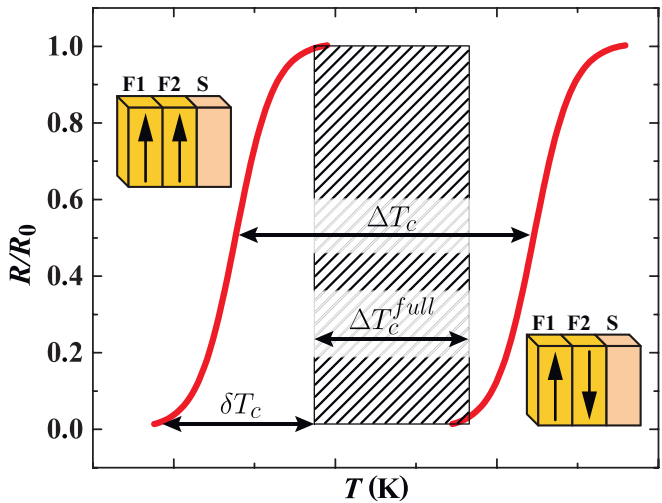}} \caption{Schematics of the operation principle of the SSV. SC transition curves (resistivity ratio $R/R_0$ as a function of $T$) with the width $\delta T_c$ for the P ($\uparrow\uparrow$) and AP ($\uparrow\downarrow$) orientations of the magnetizations of the F1 and F2 layers are scketched by solid lines. Shaded rectangle depicts the operational area of the SSV. (see the text) }
\label{fig:fig1}
\end{figure}

Fig.~1 shows schematically the operation of the SSV in its most simple version. The two SC transition curves with the width $\delta T_c$  corresponding to the
P and AP  mutual orientation of the magnetizations of the F1 and F2 layers are shifted with respect to each other by $\Delta T_c=T_c^{AP}-T_c^P$ and enclose the operational area of the SSV depicted by the shaded rectangle. Here, $T_c^{AP}$ and $T_c^P$ are the SC transition temperatures for AP and
P configurations, respectively. At a fixed temperature within this rectangle the change of the mutual orientation of
magnetizations from AP to P by an external magnetic field  yields a full
switching between the SC and the normal state of the SSV. Therefore, the width $\Delta T_c^{full}$ of the shaded area in Fig.~1 is the most important parameter of the SSV. It should be noted, that the commonly used condition for the realization of the full switching effect $\Delta T_c>\delta T_c$ is, in fact, not sufficient. The actually relevant parameter $\Delta T_c^{full}$ is always smaller than $\Delta T_c$ due to the finite value of $\delta T_c$, and this difference gets larger the larger the value of  $\delta T_c$ is.  In the first experimental realization of the full switching between the
normal and SC states in the SSV in Ref.~\cite{Leksin2010} $\Delta T_c^{full}$  was of
the order of 10\,mK, only. Thus, in order to improve the
performance of the SSV it is necessary to increase this
temperature window which was attempted in a large number of works on different constructions of SSVs (see, e.g., reviews \cite{Garifullin,Blamire,Linder} and more recent publications\cite{Zhu2010,Cheng201229, Pugach2017}).

Recently, the studies of the SSV effect shifted towards 
exploring the long-range triplet component (LRTC) \cite{Bergeret}
in the superconducting condensate generated in heterostructures with non-collinear (close to orthogonal) orientations of the magnetizations of the F1 and F2 layers \cite{Fominov} (see
also more recent articles \cite{Wu20122173,Zdravkov2013,Banerjee2014,Wang2014,Flokstra2015,Dybko201548, Lenk2016957, Voltan2016, Feng2017, Srivastava2017,Moen2017,Devizorova2017, Alidoust2018}). For example, Jara
{\it et al.} \cite{Jara} have experimentally studied the SC properties of the
CoO/Co/Cu/Co/Nb SSV structure. They obtained clear evidence for LRTC
and a good agreement between theory and experiment \cite{Wu}
(the
theory of LRTC, used in that work, was based on the numerical
solution of the microscopic Bogoliubov -- de~Gennes equations).  A full switching between the
normal and SC states assisted by the triplet contribution to the SSV
effect was observed by us in the CoO$_x$/Py1/Cu/Py2/Cu/Pb structure (Py=Ni$_{0.81}$Fe$_{0.19}$) \cite{Leksin2016}. The same result was obtained by Gu {\it et
al.} \cite{Gu,Gu2015} who studied epitaxial Ho/Nb/Ho and Dy/Nb/Dy
SSV structures. They found the magnitude of the SSV effect of $\sim$
400\,mK.

At present, practically all fundamental questions of the SSV effect seem to be well understood and it comes out that
the magnitude of the SSV effect in the constructions with elemental
ferromagnets can be hardly increased anymore calling for the use of new unconventional ferromagnetic
materials. Singh {\it et al.} \cite{Aarts2015} were the first who
successfully used a half-metallic CrO$_2$ layer as the F1
layer in the F1/Cu/F2/S structure. They obtained a giant magnitude of the
triplet SSV effect in  CrO$_2$/Cu/Ni/MoGe of
$\Delta
T_c^{trip}=T_c(\alpha=0^\circ)-T_c(\alpha=90^\circ) \geqslant 0.7$\,K, where $\alpha$ is the
angle between the cooling field and the direction of the
applied magnetic field. $\alpha=0^\circ$ corresponds to the parallel (P) and
$\alpha=90^\circ$ to the perpendicular (PP) orientation of magnetizations of
the F1 and F2 layers. Singh {\it et al.} argued that the giant
magnitude of $\Delta T_c^{trip}$ is essentially due to the half-metallic CrO$_2$
layer. For the further development of SSVs it is therefore important to verify if this conclusion generally holds for other half-metallic compounds as ferromagnetic parts of an SSV.

With this motivation, we have chosen as a test material the Heusler
alloy Co$_2$Cr$_{1-x}$Fe$_x$Al$_y$, named hereafter HA. In our previous work \cite{Kamashev2017} we found that the HA films deposited on the substrate
kept at a temperature $T_{sub} \sim 300$\,K  appear to be a
weak ferromagnet, named hereafter HA$^{RT}$. In contrast, the HA film became practically half-metallic (HA$^{hot}$)  if prepared at $T_{sub}\geqslant 600$\,K. Specifically, the degree of the spin polarization (DSP) of its conduction band reaches $70-80$\,\% \cite{Kamashev2017}.  Here, the superscripts ``RT'' and ``hot'' refer to the room and high temperature values of $T_{sub}$. In the present work using HA$^{RT}$ as a weak ferromagnet on the place of the F2 layer
in the F1/F2/S structure and HA$^{hot}$ as a ``half-metal'' on the place of F1 layer we
performed a detailed 
analysis of the SSV effect in both types of heterostructures including the appropriate theoretical description of the observed phenomena. Some preliminary data on the use of  HA$^{RT}$ and  HA$^{hot}$ in different parts of the SSV were reported by us previously in Refs.~\cite{JMMM,Kamashev2019}.

The paper is organized as follows. The choice of the design and the
preparation technique of the SSVs is described in Sect.~II. In Sect.~III, we
describe their magnetic and  transport properties. Section~IV
contains the main experimental data on the SC and SSV properties of
the samples. In Sect.~V we introduce the theoretical approach which is used
in Sect.~VI for the analysis of the obtained results. Finally, the
work is summarized in Sect.~VII. Details of the theoretical calculation of the critical temperature $T_c$ of the SSV trilayers are presented in Appendix.

\section{Samples}

\subsection{Design of the heterostructures}

In the previously studied F1/F2/S heterostructures with conventional ferromagnets we have conveniently used the
antiferromagnetic CoO$_x$ layer as a bias for the F1 layer (see, e.g., Refs.
\cite{Leksin2010,Leksin2011,Garifullin,Leksin2016,JMMM}). However, this approach does not work anymore if HA$^{hot}$ is to be used as the F1 layer. At the required for the deposition of HA$^{hot}$ substrate temperature of $T_{sub}=700$\,K,
CoO$_x$ decomposes forming ferromagnetic Co, which does not pin the magnetization of HA$^{hot}$.
In this situation the use of the natural bias
of a hard ferromagnet appears to be beneficial. Therefore, in our new structure type
the role of the F2 layer adjacent to the SC Pb layer 
is played by Ni deposited at a low temperature of the substrate $T_{sub} \sim 150$\,K. The role of the soft ferromagnet 
is played by HA$^{hot}$ which was already deposited before on the substrate held at $T_{sub} \sim 700$\,K. Hence, now not the magnetization of the F2 layer is rotated by the operating magnetic field, but that of  the F1 layer, while the F2 magnetization remains
almost fixed by ``self-bias''.

The final design of two studied structure types is depicted in Fig~2.
\begin{figure}[h]
\center{\includegraphics[width=1.0\linewidth]{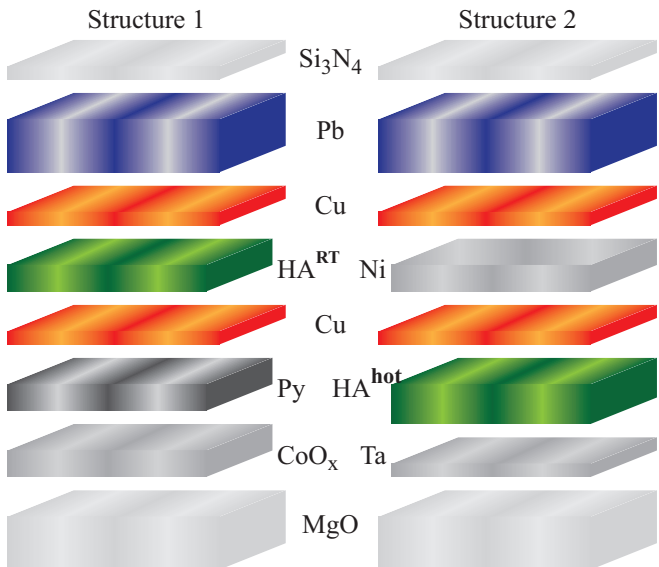}} \caption{(Color
online) Two structure types of the SSV studied in the present work. (see the text for details)}
\label{fig:fig2}
\end{figure}
In the structure Type~1 the antiferromagnetic cobalt oxide layer is used to
fix the direction of the magnetization of the Py layer (F1 layer). This allows to
rotate the magnetization direction of the weak ferromagnetic
HA$^{RT}$ layer (F2 layer) by changing the direction of the applied external magnetic
field. In turn, in the structure Type~2 the role of the ``free rotating'' layer 
is played by
the HA$^{hot}$ layer (F1 layer) and the magnetization of the Ni layer (F2 layer) is almost
fixed due to a large coercive field. The Ta layer is the seed layer for the growth of the HA$^{hot}$. The Cu layer separating the F1 and S layers prevents interdiffusion during the growth of the SSV \cite{Leksin2015}. The Cu layer between the F1 and F2 layers is used to decouple their magnetizations. Si$_3$N$_4$ is the cap protective layer.

\subsection{Preparation}

Metallic layers were grown on the high-quality single crystalline
MgO(001) substrates using classical e-gun in ultrahigh vacuum (UHV)
conditions of about $1\times10^{-8}$ mbar within a closed vacuum
cycle. The evaporation chamber has a loud-lock station allowing to
avoid vacuum breaking before the substrate load. The thickness of the
layers during the growth was controlled by a standard quartz-crystal
monitor. All materials used for evaporation had a purity of better
than 4N, i.e., the contamination level could be kept below 0.01\,at.\,\%. The substrates were fixed at a small rotating wheel on the
sample holder. After that, the sample holder was placed inside the
load-lock station. The rotating wheel system allows us to prepare a
set of samples with varied thickness at the same evacuation cycle.
The cobalt oxide layer in the structures of Type~1 was prepared in two stages.
First, the Co layer was deposited on the substrate. At the second stage,
it was transferred to the load-lock station where the Co layer was
oxidized using 2 hours' exposure at 100\,mbar of pure oxygen gas.
After the oxidation procedure, the samples were returned to the UHV
deposition chamber where the process of deposition was continued. At
the final stage, the samples were transferred to the neighboring
sputtering chamber where they were capped by a 85-nm thick
Si$_3$N$_4$ isolating protection layer to prevent the oxidation of the
top Pb layer. In order to prepare high-quality samples the
deposition rate of Pb amounted $1.0$--$1.2$\,nm/s which is very important because at lower deposition rates
the transport properties of the Pb layer do not yield the necessary
values of the SC coherence length. For other materials we used the
following deposition rates: 0.037\,nm/s for HA and 0.05\,nm/s for the Cu, Ni, Ta and Py layers. To optimize the growth of the top fragment containing the Pb layer after the deposition of the
HA we decrease $T_{sub}$ down to 150\,K and continued
the preparation process as in Ref.~\cite{Nano}. The low
$T_{sub}$ improves the smoothness of the Pb layer surface thus increasing the
magnitude of the SSV effect. The parameters of the studied samples
are presented in Table 1.

We emphasize again that the most important difference between the two structure types is
the growth temperature of the HA=Co$_2$Cr$_{1-x}$Fe$_x$Al$_y$.
According to our previous work \cite{Kamashev2017}, with  $T_{sub}=300$\,K
one obtains a weak ferromagnet HA$^{RT}$ and with  $T_{sub}\geqslant 600$\,K
an almost half-metallic HA$^{hot}$ layer with the spin polarization of the conduction
band of $70$--$80$~\% can be grown.

Finally, we mention that, as we found out before \cite{Kamashev2017}, the real stoichiometry of
HA$^{RT}$ is, actually, Co$_2$Cr$_{0.43}$Fe$_{0.36}$Al$_{0.5}$ and that of  HA$^{hot}$ is Co$_2$Cr$_{0.55}$Fe$_{0.72}$Al$_{0.62}$.
Obviously, there is a deficiency of aluminum in comparison with the ideal
Heusler composition Co$_2$Cr$_{1-x}$Fe$_{x}$Al. At the same time, in fact, this ``non-ideal''
composition demonstrates a high DSP of the order of 70\,\%. Husain {\it at al.}~\cite{Husain} have shown that the DSP increases with increasing the substrate temperature $T_{sub}$. Therefore we expect its value in our samples to be of the order of 80\,\%.

\begin{table}
\caption{ Overview of the studied samples of two types of the SSV structures (Fig.~2).\\\underline{Type~1}:
CoO$_x$(3.5nm)/Py(5nm)/Cu(4nm)/\-HA$^{RT}$($d_{HA}$)/\-Cu(1.5nm)/\-Pb(80nm); \underline{Type~2}: HA$^{hot}$(20nm)/\-Cu(4nm)/\-Ni($d_{Ni}$)/\-Cu(1.5nm)/\-Pb(105nm).}

\begin{tabular}{|c|c|c|c|}
\hline

Structure type & Sample name & $d_{HA}$ (nm) & $d_{Ni}$ (nm) \\ \hline

\multirow{3}{*}{1} & PL3481 & 0.6 & --- \\ \cline{2-4}
& PL3418 & 1 & --- \\ \cline{2-4}
& PL3416 & 4 & --- \\ \hline

\multirow{3}{*}{2} & PLAK4212 & --- & 0.9  \\ \cline{2-4}
& PLAK4214 & --- & 1.6  \\ \cline{2-4}
& PLAK4215 & --- & 2.0  \\ \cline{2-4}
& PLAK4216 & --- & 2.5  \\ \hline

\end{tabular}
\end{table}

\section{MAGNETIC AND transport CHARACTERIZATION}

\subsection{Magnetic measurements}

All samples were magnetically characterized using a standard
7T VSM SQUID magnetometer, as reported in our preliminary work \cite{JMMM}. The samples of the
structure Type~1 were cooled down
in a magnetic field of 4\,kOe applied in the sample plane and
measured at $T=10$\,K. Bearing in mind that the N\'eel temperature of
the cobalt oxide is of the order of $250$--$290$\,K, after such cooling
procedure the magnetization of the Py layer is
pinned by the anisotropy field of the antiferromagnetic
layer. For the representative sample of Type~1 (PL3481) (see Fig.~1 of our work \cite{JMMM}) the magnetization
of the free HA$^{RT}$ layer starts to decrease by decreasing the field
from $+4$\,kOe to the field of the order of $0.1$\,kOe. At
the same time, the magnetization of the Py layer is kept by the bias
CoO$_x$ layer until the magnetic field of $-2$\,kOe is reached. Thus,
in the field range between $0.1$ and $-2$ kOe the mutual orientation of
the two layers is antiparallel. While further changing the field from $-2$
to $-2.5$\,kOe the magnetization of the Py layer rotates away from the
direction of the cooling magnetic field in the direction of the applied field. Qualitatively, this kind of
the magnetic hysteresis loop is typical for all samples of the structure Type
1. The minor hysteresis loop  which arises due to reversal
of the magnetization of the F2 layer (HA$^{RT}$) has been measured in the
SSV  samples by sweeping the magnetic field $H$ from $+1$\,kOe
down to $-1$\,kOe and back to $+1$\,kOe. A typical for Type~1 samples minor loop is shown in
Fig.~1(b) in Ref.~\cite{JMMM}. The measurements were performed for the samples with
$d_{HA^{RT}}$ ranging from 0.6 to 7\,nm. It was found that the saturation
field of the HA decreases gradually (as approximately
$1/d_{HA^{RT}}$) from 1\,kOe for the sample with $d_{HA^{RT}}=0.6$\,nm down to 0.1\,kOe 
for the sample with $d_{HA^{RT}}=7$\,nm. We found a sharp decrease of the
coercive field from 1\,kOe for the sample
with $d_{HA^{RT}}=0.6$\,nm down to 0.05\,kOe for the sample with
$d_{HA^{RT}}=7$\,nm.  The height of the $M(H)$ loop due to the
reversal of the magnetization of the F1 and F2 layers is
proportional to the thickness of the respective layers. For example,
the change of the magnetization in the major loop in Fig.~1(a) in Ref.~\cite{JMMM} due
to flip of the magnetization of the pinned F1 layer is approximately
5 times larger compared to the height of the minor loop in Fig.~1(b) in Ref.~\cite{JMMM},
in agreement with the ratio of the thicknesses of the respective
layers in this representative sample.

As to the samples of structure Type~2,
one observes the onset of the saturation of the
HA$^{hot}$ magnetization at 30\,Oe. 
The magnetic response from the Ni layer is not seen due to a relatively small value of the magnetic moment of this layer. 
It should be noted that such conventional measurements 
of the magnetic hysteresis loop for the sample without HA$^{hot}$ 
by decreasing the 
magnetic field down to zero and increasing it again in the negative direction 
are different from the situation when the field vector rotates
in the plane of the sample. In the 
latter case the magnetization start 
to follow the field with a considerable delay.

\subsection{Transport properties}

The electrical resistivity was measured using a $dc$ current mode in a
standard four-terminal configuration. The temperature of the
sample was controlled with the 230\,Ohm Allen-Bradley resistor
thermometer which is particular sensitive in the temperature
range of interest. The current and voltage leads were attached to
the sample by clamping contacts. The critical temperature $T_c$ was defined as the midpoint
of the transition curve. We found that the residual resistivity
ratio $RRR=\rho(300K)/\rho(10K)$ of the studied samples lies in the
interval $10<RRR<17$. Using $\rho(300K)=21 \ \mu \Omega \cdot$cm
\cite{Kittel} we obtain $\rho_0=1.2$--$2.1 \ \mu \Omega \cdot$cm  for
the residual resistivity.  The BCS coherence length for Pb amounts
to $\xi_0 =83$\,nm \cite{Kittel} and the mean-free path of the
conduction electrons obtained using the Pippard relations
\cite{Pippard} is about $l_S=17$\,nm. The comparison of $l_S$ with
$\xi_0$ shows that $l_S\ll\xi_0$ implying the ``dirty'' limit for the
superconducting part of the system.

Thus, the residual resistivity at 4 K $\rho$(4K) was found to be
equal to 1.5 $\mu \Omega \cdot$cm. The corresponding SC coherence length reads:
\begin{equation}
\xi_S=\sqrt{\frac{\hbar D_S}{2\pi k_B T_{cS}}}.
\end{equation}
Here $T_{cS}=7.18$ K is the critical temperature of the SC Pb layer;
$D_S$ is the diffusion constant in the SC layer. For all studied samples
we obtain using Eq.~(1) $\xi_S=41$ nm.

We also tried to estimate the residual resistivity of the F2 layer
in both structure types in the thickness range corresponding to the
studied SSV samples. We cannot measure directly the partial
resistivity of each layer. The main contribution to the residual
resistivity of the F2 layers is given by the surface relaxation of
conduction electrons and, therefore, by the roughness of the F layers. Hence, we
measured resistivity of such thin layers using a single layer or
bilayer films. It turns out that the HA$^{RT}$ and Ni films grown directly
on the MgO substrate at room temperature become discontinuous at
thicknesses below 10\,nm. Therefore, we prepared a set of samples
MgO/Cu(4nm)/HA($d_{HA^{RT}}$) and MgO/Cu(4nm)/Ni($d_{Ni})$. For them, 
the quality of the HA$^{RT}$ and Ni layers was much better, and  we 
obtained $\rho_0^{Ni}\sim 40-50$\,$\mu \Omega\cdot$cm. Such a big
scatter of the resistivity values is caused by different roughness
of the layers. Possibly, the residual resistivity of the Ni layers
in the respective SSV could be different. In our analysis below we shall
use an ``optimistic'' value of the residual resistivity for the Ni layer
$\rho_0^{Ni}=40$ $\mu \Omega  \cdot$cm. Similar to
our previous work \cite{Leksin2015}, we obtain  $D_F^{Ni}=2.5$ cm$^2$/s.

The nonmagnetic coherence length in the F layers is defined as
$\xi_F=\sqrt{\hbar D_F / 2\pi k_BT_{cS}}$.
From this equation we get $\xi_F^{Ni}=6.3$ nm. As to the HA layers, the
coherence length obtained from the resistivity differs considerably
from that obtained from the fitting of the theory to the experimental
data. For the films prepared at $T_{sub}=300$\,K we obtain that the
resistivity does not depend on temperature and amounts to $\rho_F \simeq
143\ \mu \Omega \cdot$cm  (cf. 220\,$\mu \Omega \cdot$cm
in Ref.~\cite{Kudryavtsev} and 170\,$\mu \Omega \cdot$cm in Ref.~\cite{Kourov}). 
For the film prepared at $T_{sub}=700$\,K we obtain
$\rho_F\simeq 130 \ \mu \Omega \cdot$cm (cf. 330\,$\mu \Omega \cdot$cm
in Ref.~\cite{Kudryavtsev} and 170\,$\mu \Omega \cdot$cm in Ref.~\cite{Kourov}).

\subsection{Dependence of the superconducting  transition temperature
on the Pb-layer thickness}

It is important to find the optimal Pb-layer
thickness for the observation of the F/S proximity effect. In
general the Pb layer should be sufficiently thin to make the whole
SC layer sensitive to the magnetic part of the structure. The
optimal thicknesses of the Pb layer for both structure types were
determined from the $T_c(d_{Pb})$ curves for the HA$^{RT}$/Cu/Pb and
Ni/Cu/Pb trilayers measured at constant $d_{{HA}^{RT}}=12$ nm and
$d_{Ni}=5$ nm, respectively (for preliminary data see our work \cite{Kamashev2019}).

At large Pb-layer thickness, $T_c$ slowly decreases with decreasing
$d_{Pb}$. Below $d_{Pb} \sim60$\,nm for HA$^{RT}$/Cu/Pb $T_c$  and
$d_{Pb}\sim 130$\,nm for Ni/Cu/Pb the $T_c$ value starts to decrease
rapidly. Below $d_{Pb}\sim$ 30 nm and below $d_{Pb}\sim 80$ nm, in
the corresponding cases, $T_c$ is less than 1.5 K. At small thicknesses
the width of the SC transition curves $\delta T_c$ gets extremely
large, of the order of 0.4 K. Bearing in mind that the influence of
the magnetic part is stronger at small Pb-layer thickness we have
chosen $d_{Pb}= 80$ and 105\,nm as the optimal thicknesses for
the SSV structures containing HA$^{RT}$ and Ni-based F2 layers, respectively.

Moreover, the above experiment is the standard procedure for a simple estimation of the boundary parameters.
For that one measures the $T_c$ ($d_{Pb}$)-dependence at fixed ferromagnetic layer thickness
larger than the penetration depth of the Cooper pairs in the ferromagnetic layers $\xi_h$. We estimated it for Ni
as 5\,nm and obtained the critical thickness of the SC layer $d_s^{crit} \sim 79$\,nm below which superconductivity
vanishes. For the structures Type~1 we obtained $d_s^{crit} \sim 23$\,nm.

Further,  following Ref.~\cite{Fominov2002} we found the Kurpiyanov-Lukichev boundary resistance parameter \cite{Kupriyanov1988} $\gamma_{bFS}$ for the F/S interface [the exact definition is given below, in Eq. (\ref{KLgammas})].  It was approximated by $\gamma_{bFS} = 0.37$ for structures Type~1 and by $\gamma_{bFS} = 0.1$--$0.4$ for structures Type~2.

\section{Experimental Results}

\subsection{Structure Type~1}
To study the angular dependence of $T_c$ on the mutual orientation
of the magnetizations we used essentially the same protocol as
for the preparation of the magnetizations of F1 and F2 layers, which was
used for the measurement of the minor hysteresis loop: the P and AP mutual
orientations of the magnetizations of the two layers were achieved at fields $H_0=+1$ and
$-1$\,kOe, respectively. For these fields both magnetizations of the F layers are P or AP
aligned for all thicknesses of the studied HA$^{RT}$ layers.

Typically, for the samples of Type 1 the maximum magnitude of $\Delta T_c^{full}$ is achieved when the mutual orientation of the magnetizations is changed form the collinear to the orthogonal one and amounts to $\sim 0.05$\,K (see Figs.~2(c) and 3(c) in Ref.~\cite{JMMM}).
For a set of the SSV samples with different $d_{{HA}^{RT}}$ we
measured the dependence of the $T_c$ value on the angle $\alpha$ between
the direction of the cooling field and the external magnetic field, both applied in the sample plane. 
\begin{figure}[h]
\includegraphics[width=1.0\linewidth]{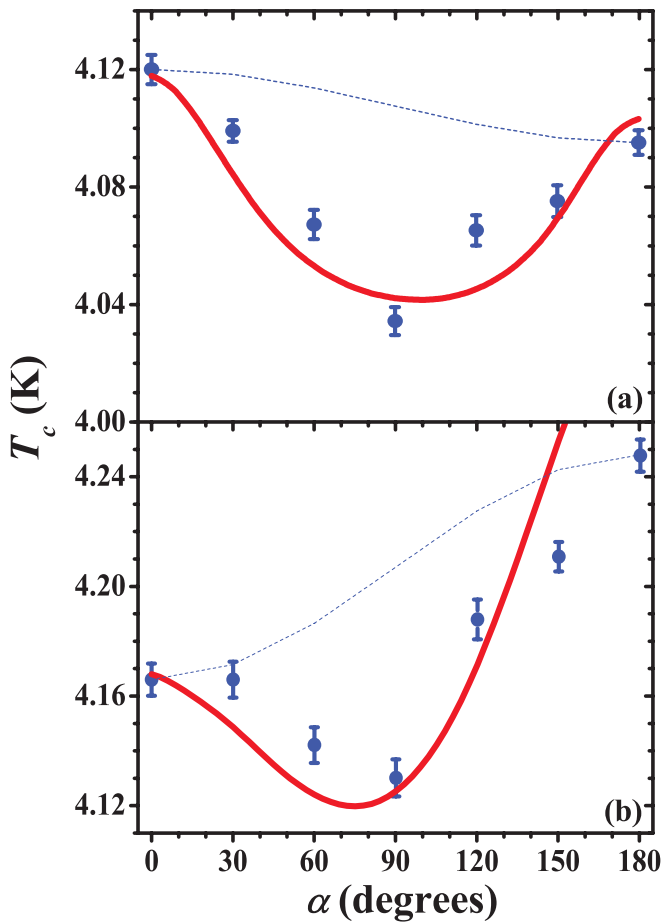}
\caption{Dependence of $T_c$ on the angle $\alpha$ between the
direction of the cooling field  used to fix the direction of the
magnetization of the Py layer and the applied magnetic field $H_0
=+1$\,kOe that rotates the magnetization of the HA$^{RT}$ layer for
samples PL3416 (a) and PL3418 (b) (experimental data points are reproduced from Ref.~\cite{JMMM}). Dashed lines depict the estimated angular dependence of the singlet SSV effect. Solid lines are the
theoretical curves with the parameters obtained in Sect.~VI (see the text for details).}
\label{fig:fig3}
\end{figure}
As can be seen from Fig.~3, while changing the mutual orientation of the
magnetizations by the gradual in-plane rotation of the magnetic field from P
($\alpha = 0^\circ)$ to the AP ($\alpha = 180^\circ)$
state, $T_c$ does not change monotonically but passes through a
minimum. According to theory \cite{Fominov}, the characteristic
minimum of the $T_c (\alpha)$ dependence is a fingerprint of LRTC.

Although the triplet component is inherent in the case of the noncollinear
magnetizations, assuming for a moment its absence one
would expect a monotonic $T_c(\alpha)$ dependence \cite{Leksin2012}. From
the symmetry arguments, $T_c(\alpha)$ should behave as $\alpha^2$ and
$(\pi -\alpha)^2$ when $\alpha$ deviates from $0$ and $\pi$,
respectively (since deviations in both limits of these
values are physically equivalent and we expect $T_c (\alpha)$ to be
an analytical function). Then one comes to a simple angular
superposition of the limiting values $T_c^P$ and $T_c^{AP}$, which reads
$T_c^{ref} (\alpha) = T_c^P \cos^2 (\alpha/2) + T_c^{AP} \sin^2
(\alpha/2)$.
This dependence is shown by the dashed lines in Figs.~3
and~6, and we treat them as reference curves. Deviation of the
actual $T_c$ from the reference curve then shows the contribution of LRTC
to $\Delta T_c$.

From Fig.~3(a) where the angular dependence of $T_c$ for the sample
PL3416 is presented, we obtain the value of the singlet SSV effect
$\Delta T_c = -25$ mK . The negative sign implies that we observe the inverse SSV effect due to the destructive interference of the Cooper pairs wave function in the heterostructure  \cite{Leksin2011}.

From Fig.~3(b) we extract the magnitude of $\Delta T_c  \sim 85$\,mK for the sample PL3418 with the positive sign corresponding to the direct SSV effect. As it can be seen from Fig.~3(b), the
difference  $T_c ^P - T_c ^{PP}$ amounts up to 100\,mK. Therefore,
the magnitude of the SSV effect with the change of the mutual orientation of the
magnetizations from AP to PP exceeds the width of the
superconducting transition curve $ \delta T_c=70$\,mK. Therefore,
there is a possibility of switching on/off the SC current fully.
As it is seen from Fig.~2(c) in Ref.~\cite{JMMM}, the full switching
$\Delta T_c^{full} \sim 50$ mK is still not very large but, nevertheless, it is 5 times
larger than in the first observation of the full SSV effect in Ref.~\cite{Leksin2010}.

\subsection{Structure Type~2}

Considering the results of the measurements of the $M(H)$ hysteresis loop of the SSV samples with the structure 
Type~2 we assumed initially that in order to manipulate the magnetization direction
of the HA layer the magnetic field of the order of 30\,Oe should be enough. We performed such experiments
and find a disappointingly small SSV effect. Surprisingly, we found that with increasing the magnetic field the
triplet contribution to the SSV effect linearly increases with magnetic field. For example,
for the sample PLAK4216 $\Delta T_c^{trip}$ increases linearly up to 0.4\,K at 2\,kOe (Fig.~4).
Notably, a similar increase of $\Delta T_c^{trip}$ was observed by Singh \textit{et al}.~\cite{Aarts2015} as well, and,
similar to us, they did not find conclusive explanation for this surprising observation.
Obviously, this field dependent effect observed by two groups on different samples is a very important
finding as it appears to be a salient feature of the new type
of SSVs and needs theoretical explanation.

Fig.~5 shows the SC transition curves for sample
PLAK4216 measured in a strong magnetic field. The shift of the curves between the P  and PP
orientations $\Delta T_c^{trip}$ 
amounts to 0.51\,K. Other samples demonstrate a smaller effect (cf. Fig.~4). 
\begin{figure}
\includegraphics[width=1.0\linewidth]{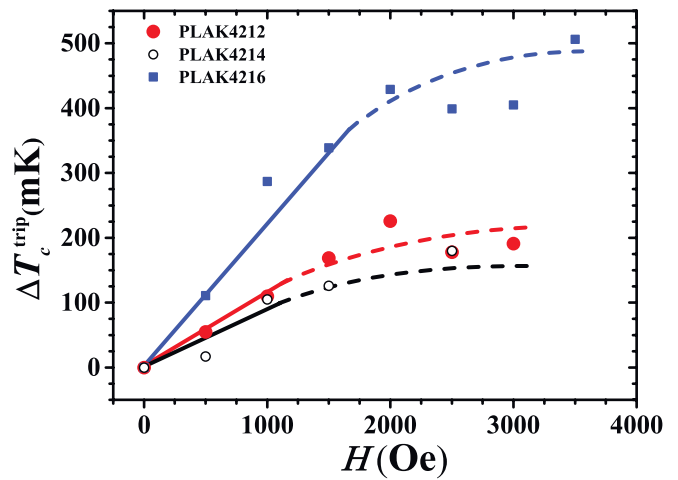}
\caption{(Color online) The magnitude of the triplet SSV effect $\Delta T_c^{trip}$ as a function of the applied
magnetic field. Lines are guides for the eye. Experimental data points for  sample PLAK4216 are reproduced from Ref.~\cite{Kamashev2019}.}
\label{fig:fig4}
\end{figure}
Fig.~6 depicts the  dependence of $T_c$ on $\alpha$ for 
two different samples. It appears qualitatively similar to the ones observed
previously in Refs.~\cite{Garifullin,Leksin2016,JMMM}, reaching a minimum near $\alpha=90^\circ$. However, the
minimum which we observe now is much deeper, suggesting that the SSV
effect is dominated by the spin polarized (triplet) Cooper pairs. Indeed, the estimated contribution of the singlet SSV effect depicted by dashed lines in Fig.~6 is practically negligible.
\begin{figure}[h]
\center{\includegraphics[width=1.0\linewidth]{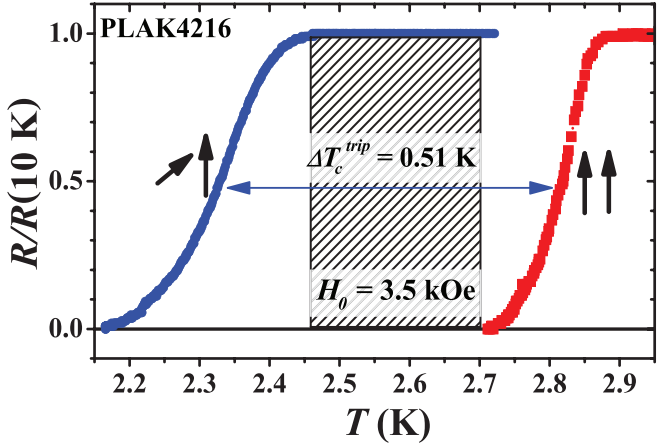}} \caption{(Color
online) Record splitting of the SC transition curves for the P  and PP  
configurations of the magnetization of the Ni layer and the applied magnetic field $H_0=3.5$\,kOe which
rotates the magnetization of the HA$^{hot}$ layer for the 
sample PLAK4216 
(Experimental $R(T)$ dependences are reproduced from Ref.~\cite{Kamashev2019}).
Shaded rectangle marks
the operational area of the SSV.}

\label{fig:fig5}
\end{figure}
\begin{figure}[h]
\includegraphics[width=1.0\linewidth]{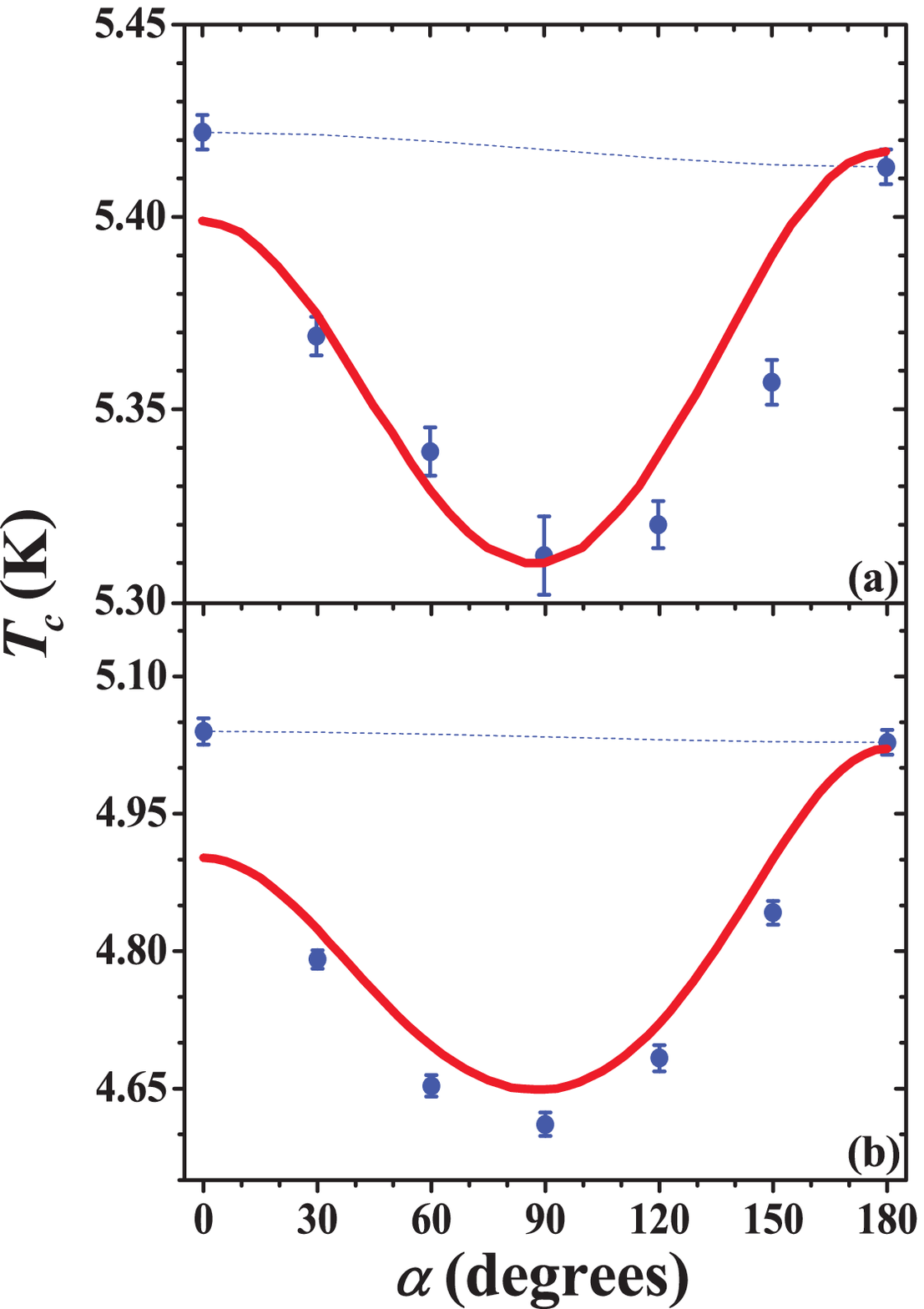}
\caption{Dependence of $T_c$ on the angle $\alpha$ between the
direction of the cooling field  used to fix the direction of the
magnetization of the Ni layer and the applied magnetic field $H_0
=1$\,kOe which rotates the magnetization of the HA$^{RT}$ layer for
samples PLAK4212 (a) and for the sample PLAK4216 (b) (experiemntal points in (b) are reproduced from Ref.~\cite{Kamashev2019}). Dashed lines 
depict the estimated angular dependence of the singlet SSV effect. Solid lines are the
theoretical curves with the parameters obtained in Sect.~VI (see the text for details).}
\label{fig:fig6}
\end{figure}
For the samples with structure Type~2 showing a large magnitude of the SSV
effect we observe an increase of $\delta T_c$ at the PP configuration
of the magnetizations (Fig.~5). In addition, the magnitude of the
triplet SSV effect $\Delta T_c^{trip}$ depends on the applied
magnetic field up to a certain value of $H_0$ which is different for different samples (Fig.~4).

\section{Theory}

The theoretical approach that we employ for the analysis of
experiment in Figs.~3 and 6, is based on the Usadel-equation technique \cite{Usadel} and generalizes the method of
Ref.~\cite{Fominov} along the lines of Ref.~\cite{Deminov2015}. It allows to consider layered
structures with different material parameters of all the layers and
arbitrary Kupriyanov-Lukichev (KL) boundary parameters
\cite{Kupriyanov1988} of all the interfaces.

Each of the two interfaces (F1/F2 and F2/S) is described by the
matching parameter $\gamma$ and the resistance parameter $\gamma_b$.
Introducing the (nonmagnetic) coherence lengths of the F layers as
\begin{equation}
\xi_{F1} = \sqrt{ \frac{\hbar D_{F1}}{2\pi k_B T_{cS}} },
\quad
\xi_{F2} = \sqrt{ \frac{\hbar D_{F2}}{2\pi k_B T_{cS}} },
\end{equation}
where $T_{cS}$ is the bulk critical temperature of the superconductor (or, equivalently, the critical temperature of an isolated S layer),
we can define the KL interface parameters as \cite{Deminov2015}
\begin{align}
\gamma_{FF} = \frac{\rho_{F2} \xi_{F2}}{\rho_{F1} \xi_{F1}},
\quad
\gamma_{bFF} = \frac{R_{bFF} \mathcal A}{\rho_{F1} \xi_{F1}},
\notag \\
\gamma_{FS} = \frac{\rho_S \xi_S}{\rho_{F2} \xi_{F2}},
\quad
\gamma_{bFS} = \frac{R_{bFS} \mathcal A}{\rho_{F2} \xi_{F2}},
\label{KLgammas}
\end{align}
where $\rho$ with a subscript is the resistivity of the corresponding layer, $\mathcal A$ is the interface area, while $R_{bFF}$ and $R_{bFS}$ are the interface resistances of the F1/F2 and F2/S interfaces, respectively.

The necessity to consider arbitrary F1/F2 interface
parameters is due to different materials of the two F layers. This
is a new theoretical ingredient, in comparison to fittings of our
previous experiments in
Refs.~\cite{Leksin2012,Leksin2015,Garifullin}.

Details of the theory are presented in Appendix.

\section{Analysis}

Theory (Sect.~V and Appendix) contains 9 fitting parameters. Our transport
measurements allow us to reduce this number down to 4.
We took $\rho_S$ and $\xi_S$ and also $\rho_0^{Ni}$ and $\xi_F$ 
as estimated in Sect.~III.B (rows 2 and 3 in Tables II and III) 
Then using Eq.~(3) we determine $\gamma_{FS}$ (rows 5 in Tables II and III) and $\gamma_{FF}$ (rows 7 in Tables II and III).
As a rough approximation
we use the theoretical results of Ref.~\cite{Fominov2002} for $d_S^{crit}$ which can be estimated in the
limit $(\gamma/\gamma_b)(d_S /\xi_S)\ll1$ as
\begin{equation} \label{dScrit}
{{d_S^{crit}}\over{\xi_S}}=2 \gamma_E
\left({{\gamma}\over{\gamma_b}}\right).
\end{equation}
Here $\gamma_E=1.78$ is the Euler constant. The value of
$\gamma_b=0$ corresponds to the fully transparent F/S interface.
$d_S ^{crit}$ is smaller for a larger value of $\gamma_b$. We use
$\rho_S \simeq 1.6\ \mu \Omega \cdot$cm, $\xi_S=41$ nm, and $\rho_F
\simeq 130\ \mu \Omega \cdot$cm. At the beginning, from Eq.~(3) we
obtain $\gamma_{FS} \simeq 0.034$ for both series of the samples. Then, taking
into account that $d_{Pb}^{crit}= 23$\,nm for the structure Type~1 and
$d_{Pb}^{crit}= 79$\,nm for the structure Type~2, we find from Eq.~(\ref{dScrit})
the values of $\gamma_{bFS} \simeq 0.2$ and $\simeq
0.65$, respectively. Using these parameters  as starting values we
have fitted the theoretical curves $T_c(d_{Pb})$. The
optimal values of the parameters obtained by the fitting are shown in Tables~II and~III. 
Obviously, the theory and experiment agree reasonably well.
\begin{table}
\label{tab:PL341} \caption{Parameters used for fitting the
experimental data on $T_c(\alpha)$  for the
CoO$_x$(3.5nm)/\-Py(5nm)/\-Cu(4nm)/\-HA($d_{HA^{RT}}$)/\-Cu(1.5nm)/\-Pb(80nm)
SSV of the structure Type~1.}

\begin{tabular}{|c|c|c|} \hline

 & PL3418 & PL3416 \\ \hline

{$d_{HA}$,\ nm} & 1 &4 \\ \hline

$\xi_S$,\ nm & 41 & 41  \\ \hline
$\xi_{F2}$,\ nm & 14 & 14 \\ \hline $\xi_{F1}$,\ nm & 7.5 & 7.5 \\
\hline
$\gamma_{FS}$ & 0.185 & 0.199  \\ \hline $\gamma_{bFS}$ & 0.37 &
0.37 \\ \hline $\gamma_{FF}$ & 1 & 1  \\ \hline $\gamma_{bFF}$ & 0
& 0  \\ \hline $h_2$,\ eV & 0.1 & 0.1 \\ \hline $h_1$,\ eV & 1 & 1 \\
\hline

\end{tabular}
\end{table}

Finally, we comment the $\gamma_{bFF} = 0$ choice for the Type I samples. We work within the Usadel equations with the Kupriyanov-Lukichev boundary conditions, and the latter contain the interface resistance $R$ as a parameter. These are model boundary conditions, and different band structures on the two sides of the interface are not taken into account within this framework. So, $R$ in the KL boundary conditions originates from a barrier at the interface. Of course, in reality a difference of the band structures can also contribute to $R$, but there is no easy way to take this effect into account. As long as we do not expect insulating barrier at the interface, we try fitting with $R=0$ (hence, $\gamma_{bFF}=0$). 
Poor fitting results in this case would tell us that either a barrier exists or that the band-structure mismatch cannot be neglected. At the same time, the result of the fitting in Figs.~3 and 6 looks satisfactory. This signifies that our initial guess of $\gamma_{bFF}=0$ was reasonable, so we go on with this value.

\subsection{Structure Type~1}

Up to now, the switching of the SC current was performed by changing
the mutual direction of the magnetizations of the F layers from AP to P
orientation (see, e.g., Fig.~1) or by combination of the singlet and
triplet SSV effect (Fig.~2 in Ref.~\cite{Leksin2016}). In both cases the full
switching between AP and PP configurations of magnetizations was obtained. It
should be noted that for the sample PL3416 the difference
$T_c^{AP}-T_c^P=60$\,mK is smaller than the difference between the P and
PP configurations which amounts to 100\,mK. Thus, the main role in the
switching is played by the triplet contribution.

\subsection{Structure Type~2}

A remarkably large separation of the SC transition curves for the P
and PP orientation of magnetizations of F1 and F2 layers in Fig.~5 yielding
the value of $\Delta T_c^{trip}$ up to 0.5\,K evidences prominent
spin-triplet superconducting correlations in our samples. Fig.~6
demonstrates that the theory correctly reproduces characteristic
features of the $T_c(\alpha)$ dependence (triplet SSV
behavior) with parameters listed in Table~III.
\begin{table}
\label{tab:PLAK421} \caption{Parameters used for fitting the
experimental data on $T_c(\alpha)$ for the
MgO/Ta(5nm)/\-HA$^{hot}$(20nm)/\-Cu(4nm)/\-Ni($d_{Ni}$)/\-Cu(1.5nm)/\-Pb(105nm) SSV of the structure Type~2.}

\begin{tabular}{|c|c|c|c|} \hline

 & PLAK4212 & PLAK4214 & PLAK4216 \\ \hline

$d_{Ni}$,\ nm &0.9 &1.6 & 2.5 \\ \hline

 $\xi_S$,\ nm & 41 & 41 & 41 \\ \hline

$\xi_{F2}$,\ nm & 6.25 & 6.25 & 6.25\\ \hline $\xi_{F1}$,\ nm & 40 &
40 & 40 \\ \hline
$\gamma_{FS}$ & 0.165 & 0.164 & 0.1  \\ \hline $\gamma_{bFS}$ & 0.4
& 0.35 & 0.1 \\ \hline $\gamma_{FF}$ & 1 & 1 & 1 \\ \hline
$\gamma_{bFF}$ & 1.5 & 1 & 0.1 \\ \hline $h_2$,\ eV & 0.03 & 0.03 &
0.03 \\ \hline $h_1$,\ eV & 0.39 & 0.39 & 0.39 \\ \hline
\end{tabular}
\end{table}

Fig.~5 shows that $\Delta T_c^{full}$ for this sample is of the order of
0.51\,K. This value is 30 times larger than it was obtained in Ref.~\cite{Leksin2010}.

As can be seen in Fig.~4, $\Delta T_c ^{trip}$ increases with
increasing the strength of the applied magnetic field. At first glance it is
surprising that $\Delta T_c^{trip}$ continues to increase well above the
saturation magnetic field for the HA$^{hot}$ layer. We suppose that
this may be caused by some magnetic inhomogeneity of the HA$^{hot}$ layer
reflected in a slight increase of its magnetization, where
more and more ``microdomains'' become gradually involved in the
formation of the total moment just as it was observed by Singh {\it
et al.} \cite{Aarts2015}.

The obtained experimental results show that as a result of the
optimal choice of materials for F layers the triplet contribution is
probably always dominant in the SSV effect. According to the
results of the present work and the data of Ref.~\cite{Aarts2015},
it appears that, indeed, a half-metallic compound is possibly the best 
presently known candidate for the material of the F1 layer in the F1/F2/S SSV. 
Its efficiency is likely related to the fact that electrons incident on the 
surface of a half metal can only
penetrate into it when they have a certain direction of spin. This concerns also
the spin polarized Cooper pairs which, depending on their specific spin orientation,
will be either reflected from the S/F interface, or easily penetrate through
it.

\section{Conclusions}

By studying the SSV  multilayers Co$_2$Cr$_{1-x}$Fe$_x$Al/Cu/Ni/Pb
whose magnetic part contains the Heusler alloy
Co$_2$Cr$_{1-x}$Fe$_x$Al with different degree of spin polarization
of the conduction band we have obtained a large SSV effect due to
the long-range triplet component of the superconducting condensate
$\Delta T_c^{trip} \sim  0.5$\,K at a moderate applied field of
3.5\,kOe as compared with the earlier work in Ref.~\cite{Aarts2015}.

In particular, finding  the most appropriate half
metal with a high degree of spin polarization of the conduction band for the F layer in the SSV
appears to be a crucial issue. At present it allows us to increase the area
of the full switching $\Delta T_c^{full}$ up to 0.3\,K which is
thirty times larger than that obtained in the first experiment
\cite{Leksin2010}. Furthermore, noting first theoretical attempts to include
the half metal into the SSV construction in
Refs.~\cite{Mironov,Halterman}, our data as well as the results by
Singh {\it et al.} \cite{Aarts2015} call for a comprehensive
quantitative theoretical treatment to obtain further insights into
exciting physics of the triplet superconducting spin valves.

\acknowledgments
Ya.V.F.\ was partially supported
by the RAS program ``Contemporary problems of low-temperature physics'',
by the RFBR (grant No.\ 19-52-50026),
and by the Basic research program of HSE.

\appendix*

\section{Theory of $T_c$ in FFS trilayers}

\begin{figure}[h]
\center{\includegraphics[width=\columnwidth]{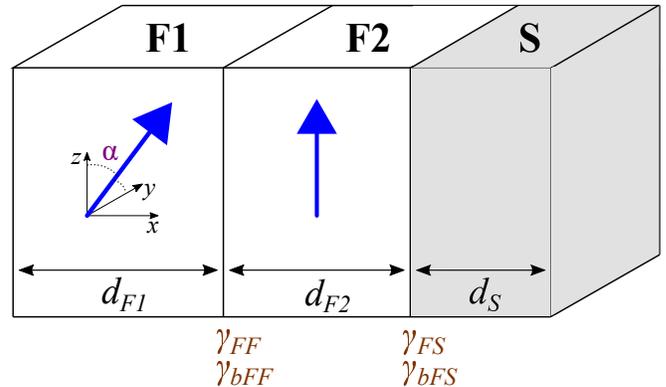}}
\caption{F1/F2/S trilayer. The F2/S interface corresponds to $x = 0$. The thick blue arrows in the F layers denote the exchange fields lying in the $(y,z)$ plane. The angle between the in-plane exchange fields $\mathbf h_1$ and $\mathbf h_2$ is $\alpha$. Each of the two interfaces (denoted as FF and FS) is characterized by two interface parameters, $\gamma$ and $\gamma_b$, see Eq.~(\ref{KLgammas}).}
\label{fig:FFS}
\end{figure}

In this Appendix, we present the theoretical method for calculating and analyzing the critical temperature of diffusive F1/F2/S trilayers (see Fig.~\ref{fig:FFS}) in the framework of the Usadel equation \cite{Usadel} and Kupriyanov-Lukichev (KL) boundary conditions \cite{Kupriyanov1988}. Within this framework, we consider the general situation assuming different exchange fields $\mathbf h_1$ and $\mathbf h_2$ in the F layers and arbitrary KL parameters $\gamma$ and $\gamma_b$ of the two interfaces (F1/F2 and F2/S). This full proximity-effect model, generalizing the model of Ref.~\cite{Fominov}, has been previously analyzed in Refs.~\cite{Deminov2015} (see also Ref. \cite{Avdeev2016453}); however, the theoretical formalism was only briefly outlined. Here, we present details of the corresponding derivation. In particular, we formulate explicit matrix equation for determining the interface function $W$. This function directly determines suppression of $T_c$ in the structure.

As shown in Fig.~\ref{fig:FFS},
the S layer is of thickness $d_S$ ($0<x<d_S$), the middle F layer is of thickness $d_{F2}$ ($-d_{F2}<x<0$),  the outer F layer is of thickness $d_{F1}$ ($-d_{F1}-d_{F2}<x<-d_{F2}$), the $x$ axis is normal to the plane of the layers. The exchange field in the middle F layer is along the $z$ direction, $\mathbf h_2=(0,0,h_2)$, while the exchange field in the outer F layer is in the $yz$ plane: $\mathbf h_1=(0, h_1\sin\alpha, h_1\cos\alpha)$.

Near $T_c$, the Usadel equations are linearized and contain only
the anomalous Green function $\check{f}$ \cite{Bergeret,IF}:
\begin{equation} \label{eq1}
\frac{\hbar D}{2} \frac{d^{2}\check{f}}{dx^{2}}-|\omega |\check{f}-\frac{i\sgn\omega}{2}
\left\{ \hat{\tau}_{0}(\mathbf{h}\hat{\boldsymbol\sigma}),\check{f}
\right\} +\Delta \hat{\tau}_{1}\hat{\sigma}_0 =0.
\end{equation}
Here, $\check{f}$ is a 4$\times$4 matrix, $\hat{\tau}_{i}$ and $\hat{\sigma}_{i}$ are the Pauli
matrices in the Nambu-Gor'kov and spin spaces, respectively, $D$ is the diffusion constant,
and $\omega = \pi k_B T_c (2n+1)$ with integer $n$ is the Matsubara energy.
The order parameter $\Delta$ is real-valued in the S layer, while in the
F layers it is zero.
The diffusion constant $D$ acquires a proper subscript (F1, F2, or S) when Eq.\ (\ref{eq1})
is applied to the corresponding layer.

The KL boundary conditions \cite{Kupriyanov1988} for the anomalous Green functions in the three layers can be written in term of the KL interface parameters (\ref{KLgammas}) as follows.
At the F1/F2 interface ($x=-d_{F2}$),
\begin{align}
\check f_{F2} &= \check f_{F1} + \gamma_{bFF} \xi_{F1} \frac{d\check f_{F1}}{dx},
\label{bcF1/F2} \\
\xi_{F2} \frac{d\check f_{F2}}{dx} &= \gamma_{FF} \xi_{F1} \frac{d\check f_{F1}}{dx}.
\end{align}
At the F2/S interface ($x=0$),
\begin{align}
\check f_S &= \check f_{F2} + \gamma_{bFS} \xi_{F2} \frac{d\check f_{F2}}{dx},
\\
\xi_S \frac{d\check f_S}{dx} &= \gamma_{FS} \xi_{F2} \frac{d\check f_{F2}}{dx}.
\label{bcF2/S}
\end{align}
At the outer surfaces of the structure (i.e., at $x=-d_{F1}-d_{F2}$ and $x=d_S$), $d\check f/dx =0$.

The anomalous Green function $\check{f}$ can be expanded into the following components:
\begin{equation} \label{eq2}
\check{f}=\hat{\tau}_{1}\left( f_{0}\hat{\sigma}_{0}+f_{3}\hat{\sigma}_{3}+f_{2}\hat{\sigma}_{2}\right) ,
\end{equation}
where $f_{0}$ is the singlet component, $f_{3}$ is the triplet with zero projection on the $z$ axis,
and $f_{2}$ is the triplet with $\pm 1$ projections on $z$ (the latter is present only
if $\alpha \neq 0,\pi$). The singlet component is even in $\omega$ (and real-valued), while the triplet ones are odd (and imaginary): $f_{0}(-\omega ) = f_{0}(\omega)$, $f_{3}(-\omega ) = -f_{3}(\omega)$, and $f_{2}(-\omega ) = -f_{2}(\omega)$,
which makes it sufficient to consider only positive Matsubara energies, $\omega >0$.

The problem of calculating $T_c$ in the FFS structure can be reduced \cite{Fominov,Deminov2015} to an effective set of equations for the
singlet component in the S layer: the set includes the self-consistency equation and the Usadel equation,
\begin{gather}
\Delta \ln \frac{T_{cS}}{T_{c}}=2\pi T_{c}\sum\limits_{\omega >0}\left(
\frac{\Delta }{\omega }-f_{0}\right) , \label{eq4} \\
\frac{\hbar D_S}{2}\frac{d^{2}f_{0}}{dx^{2}} -\omega
f_{0}+\Delta =0, \label{eq5}
\end{gather}
with the boundary conditions
\begin{equation} \label{eq6}
 \left. \xi_S \frac{df_{0}}{dx}=Wf_{0} \right|_{x=0},\qquad \left. \frac{df_{0}}{dx}=0\right|_{x=d_{S}}.
\end{equation}
This is exactly the problem for which the multi-mode solution procedure (as well as the fundamental-solution method) was developed in Refs.~\cite{Fominov2002,Fominov2}.

The difference of our current consideration from previous analysis of FFS structures \cite{Fominov} is the generalized form of $W$ in Eq.\ (\ref{eq6}). This interface function contains information about the part of the structure attached to the S layer (including, in particular, information about the magnitudes and orientations of the exchange fields and properties of the interfaces).
In addition to exact calculation of $T_c$, knowledge of the (real) interface function $W$ can be used for qualitative analysis of $T_c$ behavior as a function of system's parameters since larger $W$ implies stronger suppression of $T_c$.
So, our goal is to calculate $W$.

The Usadel equation (\ref{eq1}) generates
the following characteristic wave vectors (the exchange fields $h_1$ and $h_2$ are in energy units):
\begin{align}
&k_\omega =\sqrt{\frac{2\omega}{\hbar D_S}},
\quad
k_{\omega 1} =\sqrt{\frac{2\omega}{\hbar D_{F1}}},
\quad
k_{\omega 2} =\sqrt{\frac{2\omega}{\hbar D_{F2}}},
\notag \\
&k_{h1} =\sqrt{\frac{h_1}{\hbar D_{F1}}},\quad q_{h1} = \sqrt{k_{\omega 1}^2 + 2i k_{h1}^2},
\notag \\
&k_{h2} =\sqrt{\frac{h_2}{\hbar D_{F2}}},\quad q_{h2} = \sqrt{k_{\omega 2}^2 + 2i k_{h2}^2}.
\end{align}
Only $k_\omega$ appears in the solution for the S layer, while the F-layers' solutions are described by $k_{\omega j}$, $q_{h j}$, and $q_{h j}^*$ (where $j=1,2$ is the number of the F layer). Since the exchange energy is usually larger than the superconducting energy scale, $h_j \gg T_c$,
the $k_{\omega j}$ modes in the F layers (arising at noncollinear magnetizations) represent the \emph{long-range} triplet component \cite{Bergeret}, which plays the key role in the present study.

In the S layer,  the solution of Eq.\ (\ref{eq1}) is
\begin{equation} \label{fS}
\begin{pmatrix} f_0(x) \\ f_3(x) \\ f_2(x) \end{pmatrix} = \begin{pmatrix} f_0(x) \\ 0 \\ 0 \end{pmatrix} + \begin{pmatrix} 0 \\ A \\ B \end{pmatrix} \frac{\cosh\left( k_\omega (x-d_S) \right)}{\cosh\left( k_\omega d_S \right)},
\end{equation}
where $A$ and $B$ are purely imaginary.
The singlet component $f_0(x)$ in the S layer cannot be written explicitly,
since it is self-consistently related to the (unknown) order parameter $\Delta(x)$
by Eqs.\ (\ref{eq4})-(\ref{eq5}). Our strategy now is to obtain the effective boundary
conditions (\ref{eq6}) for $f_0(x)$, eliminating all other components in the three layers.

In the middle F2 layer,
\begin{multline}
\begin{pmatrix} f_0(x) \\ f_3(x) \\ f_2(x) \end{pmatrix} = C_1 \begin{pmatrix} 0 \\ 0 \\ 1 \end{pmatrix} \cosh\left( k_{\omega 2} x \right) + S_1 \begin{pmatrix} 0 \\ 0 \\ 1 \end{pmatrix} \sinh\left( k_{\omega 2} x \right)
\\
+ C_2 \begin{pmatrix} 1 \\ 1 \\ 0 \end{pmatrix} \cosh\left( q_{h2} x \right) + C_3 \begin{pmatrix} -1 \\ 1 \\ 0 \end{pmatrix} \cosh\left( q_{h2}^* x \right)
\\
+ S_2 \begin{pmatrix} 1 \\ 1 \\ 0 \end{pmatrix} \sinh\left( q_{h2} x \right) + S_3 \begin{pmatrix} -1 \\ 1 \\ 0 \end{pmatrix} \sinh\left( q_{h2}^* x \right) ,
\end{multline}
where $C_1$ and $S_1$ are purely imaginary, while $C_3 =-C_2^*$ and $S_3 = -S_2^*$.

Finally, in the outer F1 layer,
\begin{multline} \label{fF1}
\begin{pmatrix} f_0(x) \\ f_3(x) \\ f_2(x) \end{pmatrix}
=
E_1 \begin{pmatrix} 0 \\ -\sin\alpha \\ \cos\alpha \end{pmatrix}
\frac{\cosh\left( k_{\omega 1} (x+d_{F1}+d_{F2})\right)}{\cosh\left( k_{\omega 1} d_{F1} \right)}
\\
+ E_2 \begin{pmatrix} 1 \\ \cos\alpha \\ \sin\alpha \end{pmatrix}
\frac{\cosh\left( q_{h1} (x+d_{F1}+d_{F2})\right)}{\cosh\left( q_{h1} d_{F1} \right)}
\\
+ E_3 \begin{pmatrix} -1 \\ \cos\alpha \\ \sin\alpha \end{pmatrix}
\frac{\cosh\left( q_{h1}^* (x+d_{F1}+d_{F2})\right)}{\cosh\left( q_{h1}^* d_{F1} \right)} ,
\end{multline}
where $E_1$ is purely imaginary and $E_3 =-E_2^*$.

Altogether, Eqs.\ (\ref{bcF1/F2})-(\ref{bcF2/S}) produce 12 scalar boundary conditions at the two interfaces (F1/F2 and F2/S). We are mainly interested in
one of them, determining the derivative of the singlet component on the S side of the F2/S interface ($x=0$):
\begin{equation} \label{boundary3}
\left. \xi_S \frac{df_0}{dx} \right|_{x=0} = 2 \gamma_{FS} \xi_{F2} \Real \left( q_{h2} S_2 \right).
\end{equation}
The remaining 11 boundary conditions form a system of 11 linear equations for 11 coefficients entering
Eqs.\ (\ref{fS})-(\ref{fF1}):
\begin{equation} \label{system3}
\hat M
\begin{pmatrix}
C_1 \\
C_2 \\
C_3 \\
S_1 \\
S_2 \\
S_3 \\
E_1 \\
E_2 \\
E_3 \\
A \\
B
\end{pmatrix}
=
\begin{pmatrix}
f_0(0) \\
0 \\
0 \\
0 \\
0 \\
0 \\
0 \\
0 \\
0 \\
0 \\
0
\end{pmatrix}.
\end{equation}
The solution of this system is nonzero due to $f_0(0)$ in the right-hand side of the system. Finding
the $S_{2}$ coefficient [which is proportional to $f_{0}(0)$], we substitute it into Eq.\ (\ref{boundary3})
and thus explicitly find $W$ entering the effective boundary conditions (\ref{eq6}).


The $\hat M$ matrix (size 11$\times$11) contains only 53 nonzero elements, which are given by the following relations:
\begingroup
\allowdisplaybreaks
\begin{align}
&M_{1,2}= -M_{1,3} = M_{2,2} = M_{2,3} = M_{3,1} = -M_{2,10}
\notag \\
&\phantom{M_{1,2}} = -M_{3,11} = 1,
\notag \\
&M_{1,5}= M_{2,5} = -M_{1,6}^* = M_{2,6}^*  = \gamma_{bFS} q_{h2} \xi_{F2},
\notag \\
&M_{4,5} = M_{4,6}^* = -\gamma_{FS} q_{h2} \xi_{F2},
\notag \\
&M_{3,4}= \gamma_{bFS} k_{\omega 2} \xi_{F2},
\qquad
M_{5,4} = -\gamma_{FS} k_{\omega 2} \xi_{F2},
\notag \\
&M_{4,10}= M_{5,11} = -k_\omega \xi_S \tanh(k_\omega d_S),
\notag \\
&M_{6,2}= M_{7,2} = -M_{6,3}^* = M_{7,3}^* = \cosh(q_{h2} d_{F2}),
\notag \\
&M_{6,5}= M_{7,5} = -M_{6,6}^* = M_{7,6}^* = -\sinh(q_{h2} d_{F2}),
\notag \\
&M_{8,1}= \cosh(k_{\omega 2} d_{F2}),\quad M_{8,4} = -\sinh(k_{\omega 2} d_{F2}),
\notag \\
&M_{9,2}= M_{10,2} = -M_{9,3}^* = M_{10,3}^* = q_{h2} \xi_{F2} \sinh(q_{h2} d_{F2}),
\notag \\
&M_{9,5}= M_{10,5} = -M_{9,6}^* = M_{10,6}^* = -q_{h2} \xi_{F2} \cosh(q_{h2} d_{F2}),
\notag \\
&M_{11,1}= k_{\omega 2} \xi_{F2} \sinh(k_{\omega 2} d_{F2}),
\notag \\
&M_{11,4} = -k_{\omega 2} \xi_{F2} \cosh(k_{\omega 2} d_{F2}),
\notag \\
&M_{7,7} = -M_{8,7} \tan\alpha = [ 1 + \gamma_{bFF} k_{\omega 1} \xi_{F1} \tanh (k_{\omega 1} d_{F1}) ] \sin\alpha ,
\notag \\
&M_{10,7} = -M_{11,7} \tan\alpha = - \gamma_{FF} k_{\omega 1} \xi_{F1} \tanh (k_{\omega 1} d_{F1}) \sin\alpha ,
\notag \\
&M_{6,8} = -M_{6,9}^* = -[ 1 + \gamma_{bFF} q_{h 1} \xi_{F1} \tanh (q_{h 1} d_{F1}) ] ,
\notag \\
&M_{7,8} = M_{6,8} \cos\alpha, \quad M_{8,8} = M_{6,8} \sin\alpha,
\notag \\
&M_{7,9} = -M_{6,9} \cos\alpha, \quad M_{8,9} = -M_{6,9} \sin\alpha,
\notag \\
&M_{9,8} = -M_{9,9}^* = \gamma_{FF} q_{h 1} \xi_{F1} \tanh (q_{h 1} d_{F1}) ,
\notag \\
&M_{10,8} = M_{9,8} \cos\alpha, \quad M_{11,8} = M_{9,8} \sin\alpha,
\notag \\
&M_{10,9} = -M_{9,9} \cos\alpha, \quad M_{11,9} = -M_{9,9} \sin\alpha.
\label{M}
\end{align}
\endgroup
In the above relations, complex conjugation affects only $q_{h1}$ and $q_{h2}$, since all other parameters entering the expressions are real.
All the elements of $\hat M$ not mentioned in Eq.\ (\ref{M}) are equal to zero.

According to the procedure discussed above, the $W$ function can be found explicitly with the help of the $\hat M$ matrix.

\end{document}